\documentclass[superscriptaddress, secnumarabic, twocolumn, nobibnotes, aps, prd, showpacs]{revtex4-1}

\usepackage{CJK}
\usepackage{amssymb}
\usepackage{graphicx}
\usepackage{amsmath}
\usepackage{bm}
\usepackage{enumerate}
\usepackage{color}
\usepackage{ulem}
\usepackage{dcolumn}
\usepackage{epsfig}
\usepackage{epstopdf}
\usepackage{amsfonts}
\usepackage{mathrsfs}
\usepackage{multirow}
\usepackage{appendix}
\usepackage{braket}
\usepackage{verbatim}
\usepackage{latexsym}
\usepackage{makeidx}

\usepackage{color,xcolor}
\usepackage[bookmarks=false]{hyperref}
\hypersetup{colorlinks=true,citecolor=blue,linkcolor=blue,urlcolor=blue,pdfstartview=FitH,bookmarksopen=true}

\begin{document}

\title{Multiple-qubit Rydberg quantum logic gate via dressed-states scheme}

\author{Yucheng He}
\affiliation{School of Physics and Microelectronics, Zhengzhou University, Zhengzhou 450001, China}
\affiliation{School of Electrical Engineering, Zhengzhou University, Zhengzhou 450001, China}
\author{Jing-Xin Liu}
\affiliation{National Laboratory of Solid State Microstructures and School of Physics, Nanjing University, Nanjing 210093, China}
\affiliation{School of Physics and Microelectronics, Zhengzhou University, Zhengzhou 450001, China}
\author{F.-Q. Guo}
\affiliation{School of Physics and Microelectronics, Zhengzhou University,
Zhengzhou 450001, China}
\author{L.-L. Yan}
\email[Email: ]{llyan@zzu.edu.cn}
\affiliation{School of Physics and Microelectronics, Zhengzhou University,
Zhengzhou 450001, China}
\author{Ronghui Luo}
\affiliation{School of Physics and Microelectronics, Zhengzhou University, Zhengzhou 450001, China}
\author{Erjun Liang}
\affiliation{School of Physics and Microelectronics, Zhengzhou University,
Zhengzhou 450001, China}
\author{S.-L. Su}
\email[Email: ]{slsu@zzu.edu.cn}
\affiliation{School of Physics and Microelectronics, Zhengzhou University,
Zhengzhou 450001, China}
\author{M. Feng}
	\affiliation{School of Physics and Microelectronics, Zhengzhou University, Zhengzhou 450001, China}
	\affiliation{State Key Laboratory of Magnetic Resonance and Atomic and Molecular Physics,
Wuhan Institute of Physics and Mathematics, Chinese Academy of Sciences, Wuhan 430071, China}
\affiliation{Department of Physics, Zhejiang Normal University, Jinhua 321004, China}


\date{\today }

\begin{abstract}
We present a scheme to realize multiple-qubit quantum state transfer and quantum logic gate by combining the advantages of Vitanov-style pulses and dressed-state-based shortcut to adiabaticity (STA) in Rydberg atoms. The robustness of the scheme to spontaneous emission can be achieved by reducing the population of Rydberg excited states through the STA technology. Meanwhile, the control errors can be minimized through using the well-designed pulses. Moreover, the dressed-state method applied in the scheme makes the quantum state transfer more smoothly turned on or off with high fidelity and also faster than traditional shortcut to adiabaticity methods. By using Rydberg antiblockade (RAB) effect, the multiple-qubit Toffoli gate can be constructed under a general selection conditions of the parameters. 
\end{abstract}

\maketitle

\section{Introduction}
\label{secI}


Quantum computation shows its remarkable properties in the information processing and solving the intractable physical problems. Quantum algorithms designed for quantum computer are also predicted to overwhelm some traditional algorithm in many difficult tasks, such as, the factorization of large integers \cite{shor1997jc} and searching in the unsorted databases \cite{gover1996prl}, which will cost the traditional computation hundreds of years. Moreover, as the research forward, apart from its strong application, the information in qubits and its further topics about \textit{It from Qubit}, help us get deep insights in fundamental physics. Topics including holographic duality \cite{umemoto2018np}, quantum gravity \cite{qi2018np}, quantum chaos and dynamics \cite{choi2020prl}, make the quantum information science become a profound subject. 

As for the application aspect, building the reliable quantum gates is of vital requirement for quantum information processing since quantum algorithm is operated via quantum logic gates. In order to construct a quantum logic gate, a large variety of platforms and systems have been studied extensively, including the cavity quantum electrodynamics (CQED) \cite{ckarke2008nat}, trapped ions \cite{zoller1995prl,molmer1999prl}, optical lattices \cite{mikkelsen2007np,berezovsky2008sci}, quantum dot system \cite{bloch2008nat,trotzky2008sci}, diamond-based system \cite{lukin2007sci}, etc. Recent years, because of its long lifetime, clean platform with small decoherence rate and robust against spontaneous emission, Rydberg atoms were regarded as an ideal candidate for realizing quantum information processing.

Among Rydberg atoms \cite{RydbergAtoms}, there exists a phenomenon named Rydberg blockade which is induced by the large electric dipole among ground states and high-excited states. In blockade radius range, no more than one atom could be excited to Rydberg states. In contrast to the blockade regime, another interesting phenomenon was predicted by Ates et al. \cite{ates2007prl}, when operating the two-step excitation scheme to create an ultracold lattice gas. When the shifted energy compensated by the two-photon detuning, the so-called Rydberg antiblockade (RAB) regime will be constructed. Meanwhile, the corresponding experimental realization was presented by Amthor et al. \cite{amthor2010prl}. Controlling the detuning between the laser field and atomic transition could compensate the energy shift caused by Rydberg-Rydberg interaction (RRI) which motivates people to build the quantum logic gates based on RAB from both theories \cite{su2017pra, mitra2020pra, khazali2020prx, young2020arxiv,saffman2020pra} and experiments \cite{isenhower2010prl, zhang2010pra, wilk2010prl, maller2015pra, zeng2017prl, picken2018qst, levine2019prl, levine2018prl, graham2019prl, madjarov2020np}. A number of schemes for constructing Rydberg gates could be divided as following categories. One category is to construct the gates by using the dynamical \cite{18,19,20,21,22,24,26,mitra2020pra,28,30,31,32,34,35,36,37,38,39} and geometric \cite{41,42,43,44,45,46,47,48,49} phases. According to the method for constructing the interaction among atoms, they could be divided as blockade \cite{18,19,20,21,22,24,36,37,38,39,41,42,43,44,45,46,47,48,49}, antiblockade \cite{28,su2017pra,Su2018pra}, dipole-dipole resonant interaction\cite{31} and F\"{o}rster resonance \cite{32,34}. Moreover, as the idea for multibit C$_{k}\mbox{-}$NOT gate was presented\cite{isenhower2011qip}, heated researches based on many-body cases was triggered\cite{01,03,04,05,06,07,08,09,010,011,012,013}.

 Apart from the feasible platform, quantum information processing (QIP) requires short operating time and high fidelity, with the admissible error of gate operations below $10^{-4}$ for a reliable quantum processor. The most famous one is the stimulated Raman adiabatic passage (STIRAP) but the adiabatic conditions limit its operating time. Hence, quite a few of methods were presented to construct shortcuts \cite{muga2019rmp,AdolfoNJP}, for example, transitionless tracking algorithm \cite{berry2009jpa, chen2010prl, campo2013prl, chen2016sr, chen2011pra}, Lewis-Riesenfeld invariants theory\cite{lai1996pra,muga2009jpb}, dressed state \cite{baksic2016prl,wu2017qip}, universal SU(2) transformation \cite{huang2017pra}. Although these methods speed up the evolution process, there still exist some experimental difficulties. When designing the laser pulses of schemes, we may face problems that a perfect operation need an infinite energy gap or laser shape is not smooth enough to be turn on or off. Fortunately, Baksic et al. present the dressed-state scheme for solving the problems \cite{baksic2016prl}. Consequently, many ideas based on this method are extended successfully \cite{bukov2018prx, bukov2019prx, ribeiro2017prx, agarwal2018prl, kolbl2019prl, gagnon2017prl, nikolay2017rmp, zhou2017np, du2016nc}. Dressed-state approach, as a kind of shortcut to adiabatic scheme, can shorten the operating time but still remains the outstanding properties of adiabatic. Namely, by choosing dressed-state basis and controlling the controllable Hamiltonian $H_{c}$ properly, the unwanted off-diagonal terms in Hamiltonian of the whole system can be canceled dramatically. However, though the methods stated above are good to show their advantages, some defects are still remained since the dressed-state approach, comparing to STIRAP, has the defect of suffering from spontaneous emission especially in the intermediate state.


Here we present a combination of protocols that allows the advantages of each single method to cancel the defects of each other. For instance, in the Rydberg system, the spontaneous emission is extremely small \cite{stourm2019jpb} and can promote the effectiveness of dressed-state approach. Hence, it becomes a more promising scheme to prepare the  controlled-NOT (C-NOT) gate with the advantages of long lifetime, fast speed, high fidelity and robust against pulsing area as well as timing error. Moreover, we also visualize the process of the dressed-state method that could help people understand, utilize and optimize this protocol sufficiently.

The rest parts of this paper are organized as follow: First, in the section \ref{secII} we will give the basic model and effective model. Then, in Sec. \ref{secIII} we will discuss the specific physical process of the gate in our system. Next, we generalize our method to multiple qubits cases \cite{isenhower2011qip} in Sec. \ref{secIV}. The numerical simulation will be given in \ref{num}. Finally, the conclusions is given in Sec. \ref{conclusions}. 


\section{Two-qubit case}

\subsection*{A. The basic physical model} 
\label{secII}

\begin{figure}[htbp]
\centering
\includegraphics[scale=0.35]{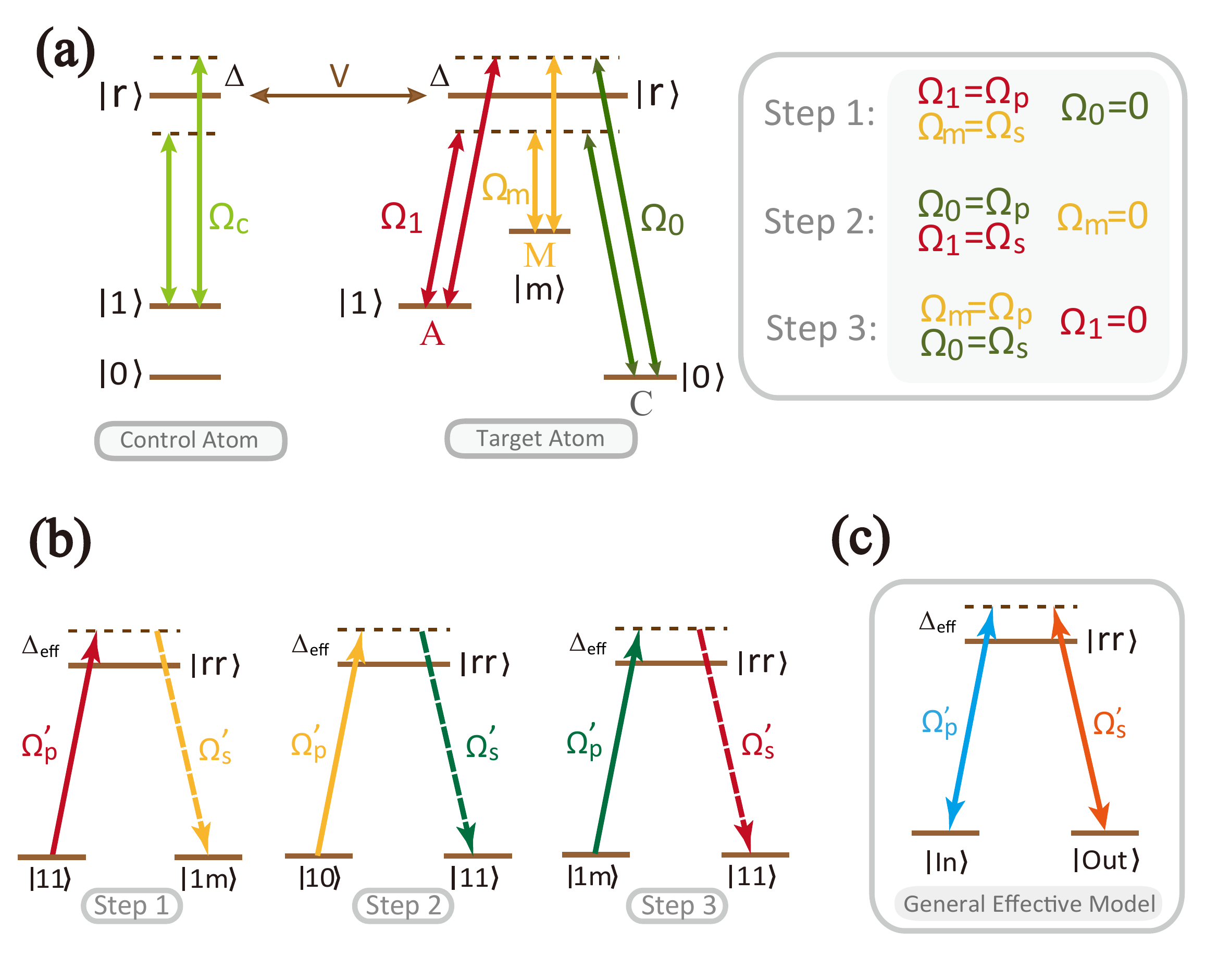}
\caption{(a) Level structure of control and target Rydberg atoms driven by the lasers to realize two-qubit quantum state transfer and quantum controlled-NOT gate. The control atom consists of two stable ground states $\ket{0}$, $\ket{1}$ and one Rydberg state $\ket{r}$ while the target atom has an additional ground state $\ket{m}$. $V$ is the Rydberg-Rydberg interaction strength between two atoms. (b) The illustration of how the three-steps process works in a effective model form. $\Delta_{\mathrm{eff}}$, $\Omega_{p}^\prime$ and $\Omega_{s}^\prime$ are the effective detuning and pulses defined in Eq. (\ref{eq:3.1}), respectively. The arrows represent the directions of population transfer among different states and the pulses have the same effective form in all the steps. (c) An uniform representation for the illustration of (b) with  $\ket{\mathrm{In}}$ (the state that population need to be transferred) and  $\ket{\mathrm{Out}}$ (the state that will accepts population transferred from $\ket{\mathrm{In}}$). }
\label{fig:1}
\end{figure}

The energy level configurations for the state transfer and quantum logic gate in two-qubit quantum system are shown in FIG.\ref{fig:1}(a). The physical system are made up of two Rydberg atoms driven by the laser pulses. The control atom has two stable ground states $\ket{0}$ and $\ket{1}$, and one Rydberg state $\ket{r}$ while the target atom has another intermediate ground state $\ket{m}$ used to temporarily store the population. Consequently, our approach to swap the populations on the states $\ket{0}$ and $\ket{1}$ of target atom when the control atom is in the ground state $\ket{1}$, divided into the following three similar steps : \textbf{i)} Add two lasers with the Rabi frequencies $\Omega_{p}$ and $\Omega_{s}$ between the Rydberg state $\ket{r}$ and two ground states $\ket{1}$ and $\ket{m}$, respectively, to let the population on $\ket{1}$ of target atom migrate to $\ket{m}$; \textbf{ii)} Add two lasers with the Rabi frequencies $\Omega_{p}$ and $\Omega_{s}$ between the Rydberg state $\ket{r}$ and two ground states $\ket{0}$ and $\ket{1}$, to make the population of $\ket{0}$ shift to $\ket{1}$; \textbf{iii)} Add two lasers with the Rabi frequencies $\Omega_{p}$ and $\Omega_{s}$ between the Rydberg state $\ket{r}$ and two ground states $\ket{m}$ and $\ket{0}$, to make the population of $\ket{m}$ transfer to $\ket{0}$. During these steps, we always keep the laser between the Rydberg state $\ket{r}$ and $\ket{1}$ of control atom keeps open. On the contrary, if the control atom is in $\ket{0}$, the state transfer between $\ket{0}$ and $\ket{1}$ of target atom will not occur. Therefor, the function of C-NOT gate between these two Rydberg atoms is realized. In the interaction picture, the Hamiltonian of this scheme can be written as $\hat{H}=\hat{H}_{1} + \hat{H}_{2} + \hat{R}$ with 
\begin{eqnarray}
&\hat{R} = V\ket{rr}\bra{rr},~~~\hat{H}_{1} = \Omega_c \cos{\Delta t} \ket{r}_{1}\bra{1} + \rm{H.c.},\cr\cr 
&\hat{H}_{2,step1} = \left( \Omega_{p} \ket{r}_{2}\bra{\mathrm{1}} + \Omega_{s} \ket{r}_{2}\bra{\mathrm{m}} + \rm{H.c.}\right) \cos{\Delta t},\cr\cr
&\hat{H}_{2,step2} = \left( \Omega_{p} \ket{r}_{2}\bra{\mathrm{0}} + \Omega_{s} \ket{r}_{2}\bra{\mathrm{1}} + \rm{H.c.}\right) \cos{\Delta t},\cr\cr
&\hat{H}_{2,step3} = \left( \Omega_{p} \ket{r}_{2}\bra{\mathrm{m}} + \Omega_{s} \ket{r}_{2}\bra{\mathrm{0}} + \rm{H.c.}\right) \cos{\Delta t},
\label{eq:2}
\end{eqnarray}
 where $V$ is the Rydberg interaction strength, $\Omega_{p} $, $\Omega_s$ and $\Omega_c$ are the corresponding Rabi frequency of lasers shown in Fig.\ref{fig:1}(a). Meanwhile, $\Delta$ represents the detuning of laser and atom transition frequency. The choices of these three pulses are flexible as long as they can meet the following conditions to structure the effective pulses and detuning: $\Omega_{p}^\prime$, $\Omega_{s}^\prime$ and $\Delta_{\mathrm{eff}}$ in Fig.\ref{fig:1}(b).
 

In large detuning condition $\Delta \gg \Omega_{c,p,s}$ and Rydberg blockade condition $V \gg \Omega_{c,p,s}$, the dynamics of system can be described by the effective Hamiltonian
\begin{equation}
\begin{aligned}
 H_{\mathrm{eff,step 1}}&= \left(\Omega_{p}^\prime \ket{\mathrm{11}}\bra{rr} + \Omega_{s}^\prime \ket{\mathrm{1m}}\bra{rr} + \rm{H.c.}\right) \\
 &+ \Delta_{\mathrm{eff}} \ket{rr}\bra{rr},\\
  H_{\mathrm{eff,step 2}}&= \left(\Omega_{p}^\prime \ket{\mathrm{10}}\bra{rr} + \Omega_{s}^\prime \ket{\mathrm{11}}\bra{rr} + \rm{H.c.}\right) \\
 & + \Delta_{\mathrm{eff}} \ket{rr}\bra{rr},\\
  H_{\mathrm{eff,step 3}}&= \left(\Omega_{p}^\prime \ket{\mathrm{1m}}\bra{rr} + \Omega_{s}^\prime \ket{\mathrm{11}}\bra{rr} + \rm{H.c.}\right) \\
 & + \Delta_{\mathrm{eff}} \ket{rr}\bra{rr},\\
\end{aligned}
\label{eq:3.1}
\end{equation}
For simplicity, we use an uniform representation, as depicted in FIG.\ref{fig:1}(c), to describe Eq. (\ref{eq:3.1}) as 
\begin{equation}
\begin{aligned}
 H_{\mathrm{eff}}&= \left(\Omega_{p}^\prime \ket{\mathrm{In}}\bra{rr} + \Omega_{s}^\prime \ket{\mathrm{Out}}\bra{rr} + \rm{H.c.}\right) \\
 & + \Delta_{\mathrm{eff}} \ket{rr}\bra{rr},
\end{aligned}
\label{eq:3}
\end{equation}
with 
\begin{equation}
\Omega_{p,s}^\prime = \frac{\Omega_c \Omega_{p,s}}{2\Delta}, \quad \Delta_{\mathrm{eff}} = V - 2\Delta + \frac{\Omega_c^{2} + \Omega_{s}^{2} + \Omega_{p}^{2}}{3\Delta},
\label{eq:3.1}
\end{equation} 
and $\ket{\mathrm{In}}$($\ket{\mathrm{Out}}$) denoting the specific two-atom states ~$\{|10\rangle,~|1m\rangle,~|11\rangle \}$ who lost (gain) population in the corresponding step. The distance between optically trapped Rydberg atoms are always considered invariant. Consequently, the distance-based Rydberg-Rydberg interaction strength $V$ will keep constant in the scheme. Therefore, by choosing appropriate parameters satisfying $\Delta_{\mathrm{eff}} = 0$, the term containing $\ket{rr}\bra{rr}$ can be dynamically eliminated and the effective model can be turned to a resonant model. Adjusting $\Delta$ and $\Omega_c$ to keep their ratio as a constant $\alpha = \Omega_c/\Delta \ll 1$,the anti-blockade condition can be obtained by the solutions of quadratic equation $(6 - \alpha^2) \Delta^2 - 3 V \Delta - (\Omega_s^2 + \Omega_p^2) = 0$. Then, the C-NOT gate is constructed through these effective Hamiltonian.

\subsection*{B. Path design via the auxiliary dressed-state basis}
\label{secIII}

\begin{figure*}[htbp]
\centering
\includegraphics[scale=0.35]{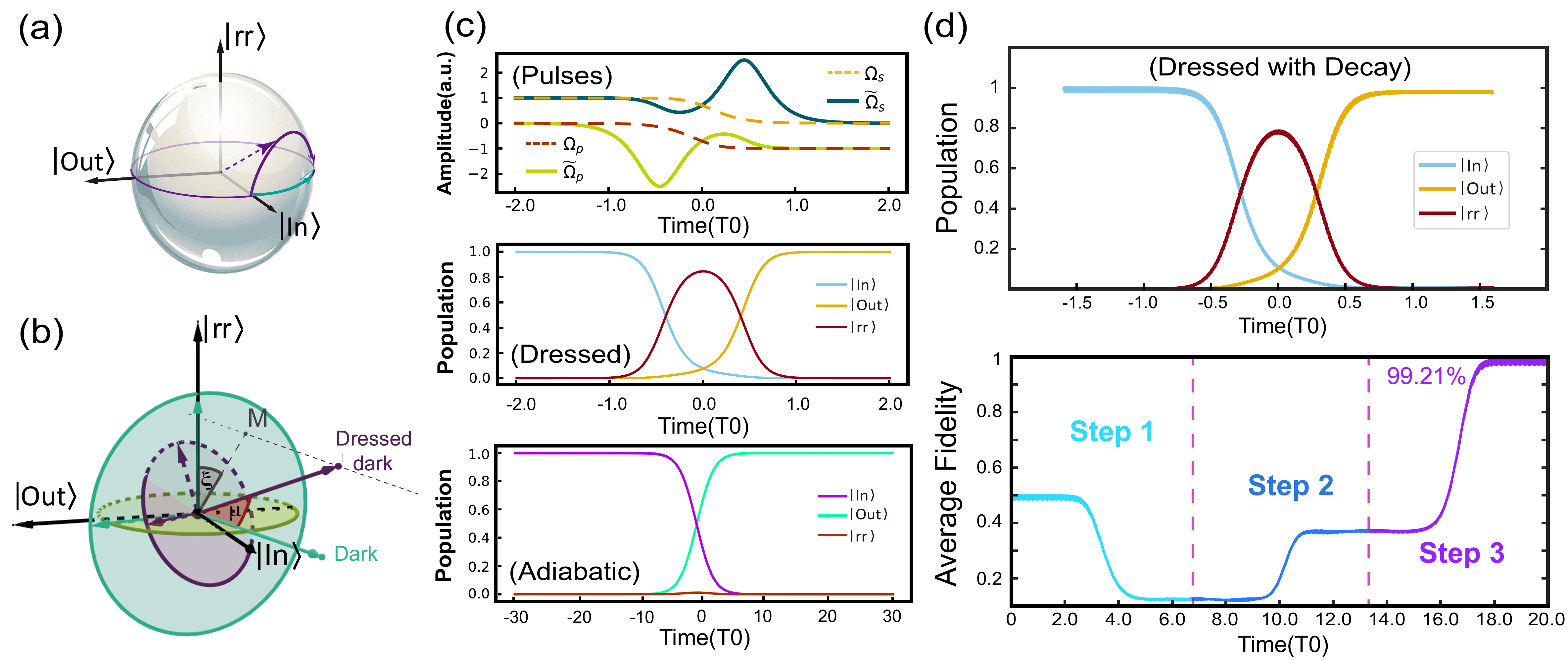}
\caption{\textbf{(a)} Consider a unit sphere with a vector pointed from the center to the surface, which represents the specific state (also noted as dressed-dark state) in a Hilbert space. The black coordinates $\ket{\rm{In}},\ket{rr},\ket{\rm{Out}}$ refer to three independent states, respectively.
The square of its projection along each axis $a^{2}, b^{2},c^{2}$ refer to the population located in corresponding states ${|\rm In\rangle,|rr\rangle,|\rm Out\rangle}$ that satisfy $\sqrt{a^{2}+b^{2}+c^{2}}=1$. The purple trace shows the path of the specified state in our method and the green trace as a contrast shows the path of traditionally adiabatic methods. Inside the sphere, the structure was shown in \textbf{(b)}. The purple coordinate system ("Dressed Dark") denotes the picture in our method while the green one ("Dark") is the picture of traditional adiabatic method. The Euler angles $\mu$ and $\xi$ are corresponding to the rotation angle and the precession angle between these two coordinates. From the geometry relations, we can establish the connection between population and Euler angles $\mu$, $\xi$ (Eq. (\ref{eq:14})). Enlarging $\xi$ to $\pi$/2, the "Dressed Dark" will be reduced to the traditional adiabatic method. \textbf{(c)} Here we choose a unit time scale T0. \textbf{Top:} The laser shape of adiabatic (dashed) and dressed (solid) methods. \textbf{Middle:} Population transition along dressed-dark state without any phase in one step. Parameters used here are: $\Omega_c/2\pi = \Omega/2\pi = 30$ MHz, $\Delta = 15\Omega$, $\tau = 0.2/\Omega$. \textbf{Bottom:} Population transition in adiabatic method. Parameters chosen:$\Omega_c/2\pi = \Omega/2\pi = 30$ MHz, $\Delta = 15\Omega$, $\tau = 3/\Omega$. \textbf{(d)} \textbf{Top:} The state transfer under decay $\gamma/2\pi = 1$ kHz.The reset of the parameters remain the same. \textbf{Bottom:}The average fidelity in two-qubit case, which can up to 99.21\%.}
\label{fig:2}
\end{figure*}

In the efficient $\Lambda$-type three-level model consisted of the Hamiltonian in Eq. (\ref{eq:3}),  we can depict, under anti-block condition $\Delta_{\mathrm{eff}} = 0$ in Eq. (\ref{eq:3.1}), a time-dependent Hamiltonian as
\begin{equation}
\hat{H}_{\mathrm{ad}}(t) = \Omega(t) \hat{M}_z + \dot{\theta}(t) \hat{M}_y,
\label{eq:6}
\end{equation}
where $\Omega(t) = \sqrt{(\Omega_s^\prime)^2 + (\Omega_p^\prime)^2}$, $\theta(t) = - \arctan{ \left( \Omega_p^\prime / \Omega_s^\prime \right)}$, and the operators
\begin{equation}
\begin{array}{cl}
\hat{M}_z &= \ket{\varphi_+}\bra{\varphi_+} - \ket{\varphi_-}\bra{\varphi_-}, \\
\hat{M}_x &= \left( \ket{\varphi_-} - \ket{\varphi_+} \right) \bra{\varphi_0}/\sqrt{2} + \rm{H.c.}, \\
\hat{M}_y &= i\left( \ket{\varphi_-} + \ket{\varphi_+} \right) \bra{\varphi_0} /\sqrt{2} + \rm{H.c.}, \\
\end{array}
\label{eq:7}
\end{equation}
are spin-1 matrices obeying the commutation relation $[\hat{M}_i,\hat{M}_j] = i\varepsilon^{ijk} \hat{M}_k$ with $\varepsilon$ denoting the anti-symmetric tensor under the adiabatic basis $\ket{\varphi_{a}} $ with $a=\pm$ and $0$ corresponding to the ``Bright" states and ``Dark" state in the effective three-level system. The off-diagonal term of $H_{\textrm{ad}}$ will increases as the operating time $\tau$ decreases. Supposing the unitary operator $\hat{U}_{\mathrm{ad}}$ connecting the original site  $\{\ket{\mathrm{In}},\ket{\mathrm{Out}},\ket{rr}\}$ with the adiabatic states $\{\ket{\varphi_{\pm,0}}\}$, it can be obtained as 
\begin{equation}
\hat{U}_{\mathrm{ad}}(t) = \frac{1}{\sqrt{2}}\left( 
\begin{matrix}
 \sin{\theta(t)} & -1 & -\cos{\theta(t)} \\
 \sqrt{2}\cos{\theta(t)} & 0 & \sqrt{2}\sin{\theta(t)} \\
 \sin{\theta(t)} & 1 & -\cos{\theta(t)} \\
\end{matrix}
\right)
\label{eq:8}
\end{equation} 
The dressed-state method is to choose a new set of dressed state $\{\ket{\phi_{d} (t)} \}$ ($d=\pm,0$ in the three-level model) to describe the Hamiltonian and cancels its off-diagonal elements in new picture that we can define an unitary operator transforming from the adiabatic basis to dressed-state basis as $V(t) = \ket{\phi_{d}(t)} \bra{\varphi_{a}}$. In order to use this method to process quantum information, an additional control Hamiltonian $H_{c}$ should satisfy:
\begin{equation}
\begin{array}{c}
\bra{\phi_{d}}H_{\mathrm{new}}\ket{\phi_{k}} = 0, d \neq k \\
V(t_i)\ket{\varphi_{a}} = V(t_f)\ket{\varphi_{a}} = \ket{\varphi_{a}} \\
\end{array}
\label{eq:4}
\end{equation}
where $H_{\mathrm{new}}$ denotes the Hamiltonian of the system after transferred to dressed-state frame:
\begin{equation}
\hat{H}_{\mathrm{new}} = \hat{V} \hat{H}_{\mathrm{ad}} \hat{V}^\dag + \hat{V} \hat{U}_{\mathrm{ad}} \hat{H}_{\mathrm{c}} \hat{U}^\dag_{\mathrm{ad}} \hat{V}^\dag + i \frac{d\hat{V}}{dt} \hat{V}^\dag.
\label{eq:5}
\end{equation}
The additional control field to modify the evolution path, is defined as
\begin{equation}
\begin{aligned}
  \hat{H}_{c}(t)=\hat{U}_{ad}^\dag(t) \left( g_{x}(t)\hat{M}_{x}+g_{z}(t)\hat{M}_{z} \right) \hat{U}_{ad}(t)
\end{aligned}
\label{eq:10}
\end{equation}
which does not directly couple $\ket{\mathrm{In}}$ and $\ket{\mathrm{Out}}$ but cancels the off-diagnoal term of $H_{\mathrm{new}}$ by controlling the parameters $g_{x}$ and $g_{z}$ satisfying
\begin{equation}
\begin{array}{cl}
g_{x}(t) &= \frac{\dot{\mu}}{\cos\xi} -\dot{\theta} \tan\xi,\\
g_{z}(t) &= -\Omega + \dot{\xi} +\frac{\dot{\mu}\sin\xi-\dot{\theta}}{\tan\mu \cos\xi}.\\
\end{array}
\label{eq:11}
\end{equation} 
Furthermore, the unitary operator $\hat{V}$ can be, using the Eular angle $\mu$, $\xi$ and $\eta$, rewritten as
\begin{equation}
\hat{V}(t)=\exp \left[i\eta(t)\hat{M}_{z}\right] \exp\left[ i\mu(t)\hat{M}_{x} \right] \exp\left[ i\xi(t)\hat{M}_{z}\right], \notag \label{eq:9}
\end{equation}
where $\eta(t)$ only changes the phase of off-diagonal element [see Appendix \ref{appendix}] and we set $\eta(t) = 0$ for simple analysis. Eventually, the Hamiltonian comes into
\begin{equation}
\hat{H}_{\mathrm{new}} = -\frac{\dot{\theta}+\dot{\xi}+\dot{\mu}\sin\xi}{\sin\mu \cos\xi}\hat{\sigma}_{z}.
\label{eq:12}
\end{equation}
where the notation $\hat{\sigma}_{z}\equiv\ket{\phi_+}\bra{\phi_+} - \ket{\phi_-}\bra{\phi_-}$.
Thus, the evolution operator in the original basis space $\{\ket{\mathrm{In}},\ket{\mathrm{Out}},\ket{rr}\}$ can be written as 
\begin{equation}
\hat{U} = U_{\mathrm{ad}}^\dag(t_f) \hat{V}^\dag (t_f) \exp{\left( -i \int_{t_i}^{t_f} \mathrm{d} \tau \hat{H}_{\mathrm{new}} \right)} \hat{V} (t_i) \hat{U}_{\mathrm{ad}}(t_i). \notag
\label{eq:13}
\end{equation}
Supposing the initial (final) state $\ket{\mathrm{In}}$ ($\ket{\mathrm{Out}}$) corresponding to the eigenstate $\ket{\varphi(t_i)}$($\ket{\varphi(t_f)}$) of system, we can select the evolution path along the dressed-dark state $\ket{\phi_0(t)}$ to accelerate the evolution process. The population of time-dependent state on $\ket{r}$ can be directly deduced from Eq. (\ref{eq:12}) or the geometric relation via the projection of state vector on $\ket{rr}$ axis of Fig.\ref{fig:2}(b) through following simple relation:
\begin{equation}
\begin{aligned}
|\braket{\psi(t)|rr}|^{2} =\sin^{2}\mu(t)\cos^{2}\xi(t).\end{aligned}
\label{eq:14}
\end{equation}
One can also simply obtain that the population $P_{\mathrm{In},\mathrm{Out}}$ on the states $\mathrm{In}$ and $\mathrm{Out}$ are corresponding to
\begin{equation}
\begin{aligned}
P_{\mathrm{In}} =[\cos\theta(t)\cos\mu(t)+\sin\theta(t)\sin\mu(t)\sin\xi(t)]^{2},\cr
P_{\mathrm{Out}} =[\sin\theta(t)\cos\mu(t)-\cos\theta(t)\sin\mu(t)\sin\xi(t)]^{2}.
\end{aligned} \notag
\label{eq:14}
\end{equation}

To keep the process smoothly turn on and off, we use Vitanov-style pulses, which can create a higher fidelity than a traditional Gaussian pulses\cite{vitanov2009pra}, to select the pulse shapes as 
\begin{equation}
\Omega_{p}^\prime(t)=-\widetilde{\Omega} \mathrm{sin}\widetilde{\theta}(t), \Omega_{s}^\prime(t)=\widetilde{\Omega} \mathrm{cos}\widetilde{\theta}(t),
\label{eq:15.a}
\end{equation}
where the amplitude of the pulses $\widetilde{\Omega}$ is a constant and the time-dependent $\widetilde{\theta}(t)$  is given by 
\begin{equation}
\widetilde{\theta}(t)=\frac{\pi}{2}\frac{1}{1+\exp{\left( -\frac{t}{\tau} \right)}}
\label{eq:15}
\end{equation}
where $\tau$ describes the smoothness of pulse shape. After being dressed, the relevant parameters should be rewritten as
\begin{equation}
\begin{aligned}
&\theta(t) = \widetilde{\theta}(t) - \arctan \left(\frac{g_{x}(t)}{\widetilde{\Omega} + g_{z}(t)}\right), \\
&\Omega(t)=\sqrt{\left[\widetilde{\Omega}+g_{z}(t)\right]^{2} + g_{x}^{2}(t)}.
\label{eq:16}
\end{aligned}
\end{equation}
The simplest non-trivial choice of the dressed-state basis \cite{baksic2016prl} is setting $\xi(t) = 0$ and $ \mu(t) = -\arctan{(\dot{\theta}/\Omega )}$ which produce
\begin{equation}
\begin{array}{ccl}
g_{x}(t) &= \dot{\mu}(t),~~g_{z}(t) &= 0,~~H_{\mathrm{new}} = -\frac{\dot{\theta}}{\sin\mu}\hat{\sigma}_{z}.
\label{eq:17}
\end{array}
\end{equation}
For each period, the population transfer thoroughly and the gate fidelity is taken as $ F = (\mathrm{tr} \sqrt{\hat{\rho}^{1/2}\hat{\sigma}\hat{\rho}^{1/2}})^{2}$. $\hat{\sigma}$ correspond to the density matrix of the current state, while $\hat{\rho}$ correspond to the expected one. In Fig. \ref{fig:2}(c, Top) we show the laser pulse before (dashed line) and after modifying the evolution path and the dynamical evolution of two-atom anti-blockade model is plotted in Fig. \ref{fig:2}(c, Middle). A small $\tau$ means the system evolves drastically and also need a strong enough modify field to realize the dressed-state method which may break the large detuning condition. On the other hand, a large $\tau$ may also prolong the operating time which corresponds to a case of adiabatic evolution (see Fig. \ref{fig:2}(b, Bottom)). By selecting appropriate parameters, the operations of state-transfer without introducing any phase can create a C-NOT gate with fidelity more than $99.21\%$.

\section{Multiple qubits case via many-body anti-blockade}
\label{secIV}

\begin{figure}[htbp]
\includegraphics[scale=0.3]{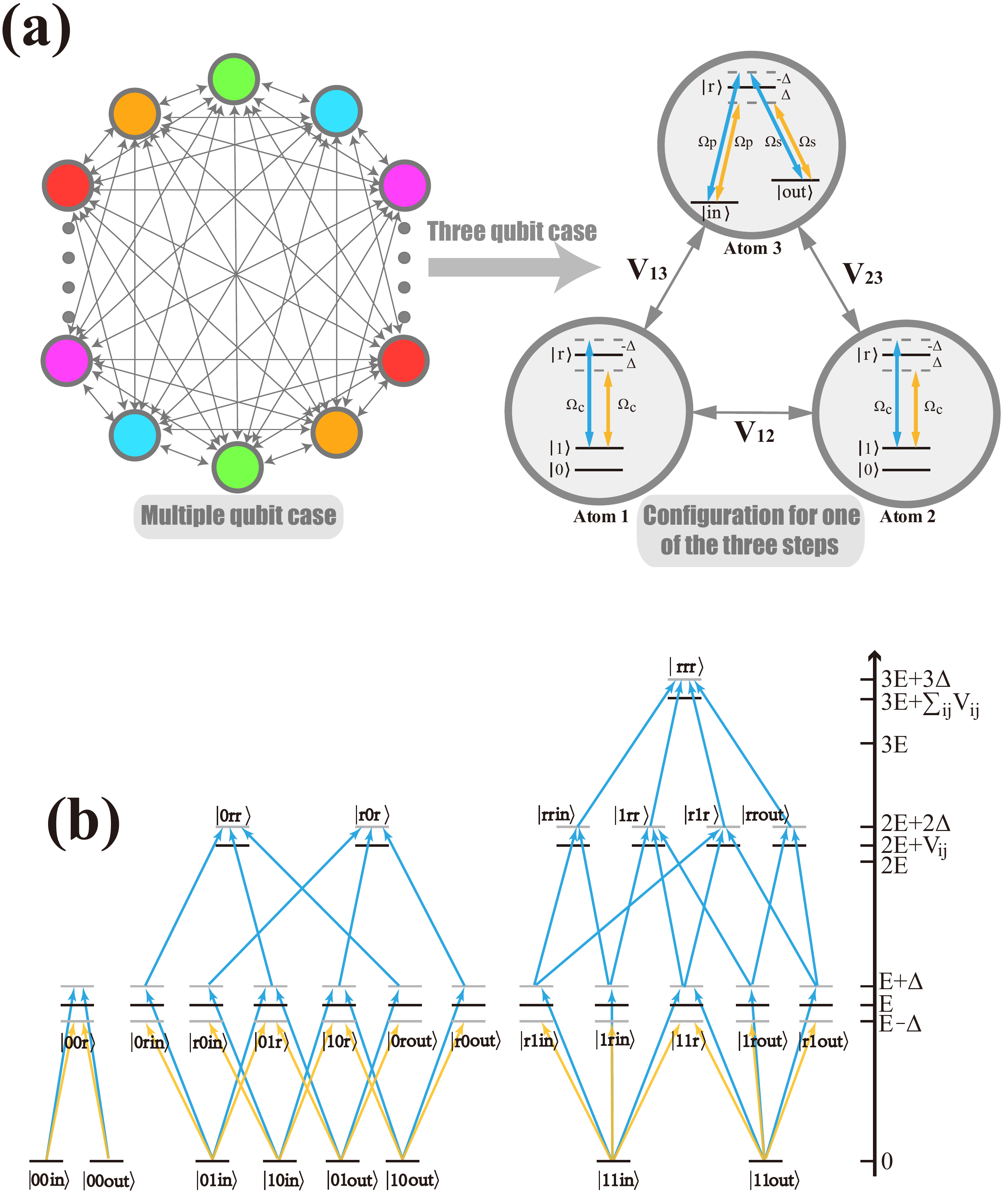}
\caption{(a) The general configuration in n-qubit case (Left). A special configuration of three Rydberg atoms (Right) in \textbf{any single steps, among all the three steps}. Atom 1 and atom 2 are control qubits, while atom 3 is the target atom. Each control atom is a three level system with two ground state $|0\rangle,|1\rangle$ and one high-layed Rydberg state $\ket{r}$ while for a target atom, there are three ground state $\ket{0}, \ket{1}, \ket{m}$ and a Rydberg state $\ket{r}$. However, for simplicity, just as we did in two-bit case, the corresponding ground states in the target atom was noted as $\ket{in}$ and $\ket{out}$ in each step. Meanwhile, $V_{12}, V_{13}, V_{23}$ among the atoms refer to the interaction strength between different atoms respectively. (b)The energy diagram and dynamic processes for a single step. }
\label{fig:3}
\end{figure}

For the multiple qubits case, we still take the three-steps process as the two-qubit case. The general form of effective Hamiltonian in $n$-body case for a specific step can be written as
\begin{equation}
\begin{aligned}
\mathcal{H}_{\mathrm{eff}}&=\frac{A^{n}_{n}\Omega^{n-1}_{c}\Omega_{p}}{2^{n}\Delta^{n-1}}\ket{rr...r}\bra{11..1\mathrm{in}}+ \Delta_{\mathrm{eff}}\ket{rr...r}\bra{rr...r}\\
&+\frac{A^{n}_{n}\Omega^{n-1}_{c}\Omega_{s}}{2^{n}\Delta^{n-1}}\ket{rr...r}\bra{11..1\mathrm{out}} + \rm{H.c.}. \notag
\label{eq:23}
\end{aligned}
\end{equation}
As a special case of $n$-qubit case (Fig. \ref{fig:3}(a, Left)) \cite{Xing2020pra,Su2018pra}, the three-qubit  interaction between different atoms and their energy-level diagram in a specific step are plotted in Fig.\ref{fig:3}(a, Right) and (b), respectively. Then, the Hamiltonian in three qubits case could be written as  $\hat{H}=\hat{H}_{\mathrm{atom}}+\hat{H}_{I}$ with
\begin{equation}
\begin{array}{cl}
H_{\mathrm{atom}}(t) & = ( \Omega_c \ket{r}_{1}\bra{1} + \Omega_c \ket{r}_{2}\bra{1} + \Omega_{p} \ket{r}_{3}\bra{\mathrm{in}} \\
& + \Omega_{s}\ket{r}_{3}\bra{\mathrm{out}} + \rm{H.c.} )  \cos(\Delta t),
\label{eq:18} \notag
\end{array}  
\end{equation}
and interaction terms
\begin{equation}
\begin{aligned}
H_{I} &= V_{12}\ket{rr}\bra{rr} \otimes I + V_{23}~I \otimes \ket{rr}\bra{rr} \\
& + V_{13}\ket{r}\bra{r} \otimes I \otimes \ket{r}\bra{r}. \notag
\label{eq:19}
\end{aligned}  
\end{equation}
Then the general effective Hamiltonian for one of the three steps can be calculated as
\begin{equation}
\begin{aligned}
\mathcal{H}_{\mathrm{eff}}&=\Omega^{(p)}_{\mathrm{eff}}\ket{rrr}\bra{11\mathrm{in}} + \Omega^{(s)}_{\mathrm{eff}}\ket{rrr}\bra{11\mathrm{out}} + \rm{H.c.}\\
& + \Delta_{\mathrm{eff}}\ket{rrr}\bra{rrr}.
\label{eq:20}
\end{aligned}
\end{equation}
where
\begin{equation}
\begin{array}{c}
\Omega^{(p)}_{\mathrm{eff}} = \frac{ 3\Omega_{c}^{2}\Omega_{p}}{4\Delta^{2} },~~\Omega^{(s)}_{\mathrm{eff}} = \frac{3\Omega_{c}^{2}\Omega_{s}}{4\Delta^{2}},\\
\Delta_{\mathrm{eff}} = V_{12} + V_{23} + V_{13} - 3\Delta + \frac{(2\Omega_c^{2}+\Omega^{2}_{p}+\Omega^{2}_{s})}{3\Delta}.
\label{eq:21} 
\end{array}
\end{equation}
In the Simplified process, the large detuning condition $\Delta \gg \max \{\Omega,\Omega_{p},\Omega_{s} \}$ should be satisfied. Since the interaction term $\ket{rrr}\bra{rrr}$ only depends on the sum of $V_{12}, V_{13}, V_{23}$, this gate can be trapped in any desired shape. Here we use the same dressed-state method in \ref{secII} for this effective three level model and the long lifetime of Rydberg state can help to suppress the occurrence of error. By choosing appropriate parameters to satisfy RAB condition
\begin{equation}
(9-2\alpha^2)\Delta^2 - 3(V_{12} + V_{23} + V_{13})\Delta - (\Omega_p^2 + \Omega_s^2) = 0,
\label{eq:22}
\end{equation}
we can observe the population transfer by using auxiliary basis method (see Fig. \ref{fig:4} (a, Top)) while when $\tau$ is chosen as a large number, the outcome corresponds to the adiabatic case Fig. \ref{fig:4} (a, Middle).   

\section{The numerical simulations of protocol}

\label{num}
\begin{figure}[htbp]
\centering
\includegraphics[scale=0.31]{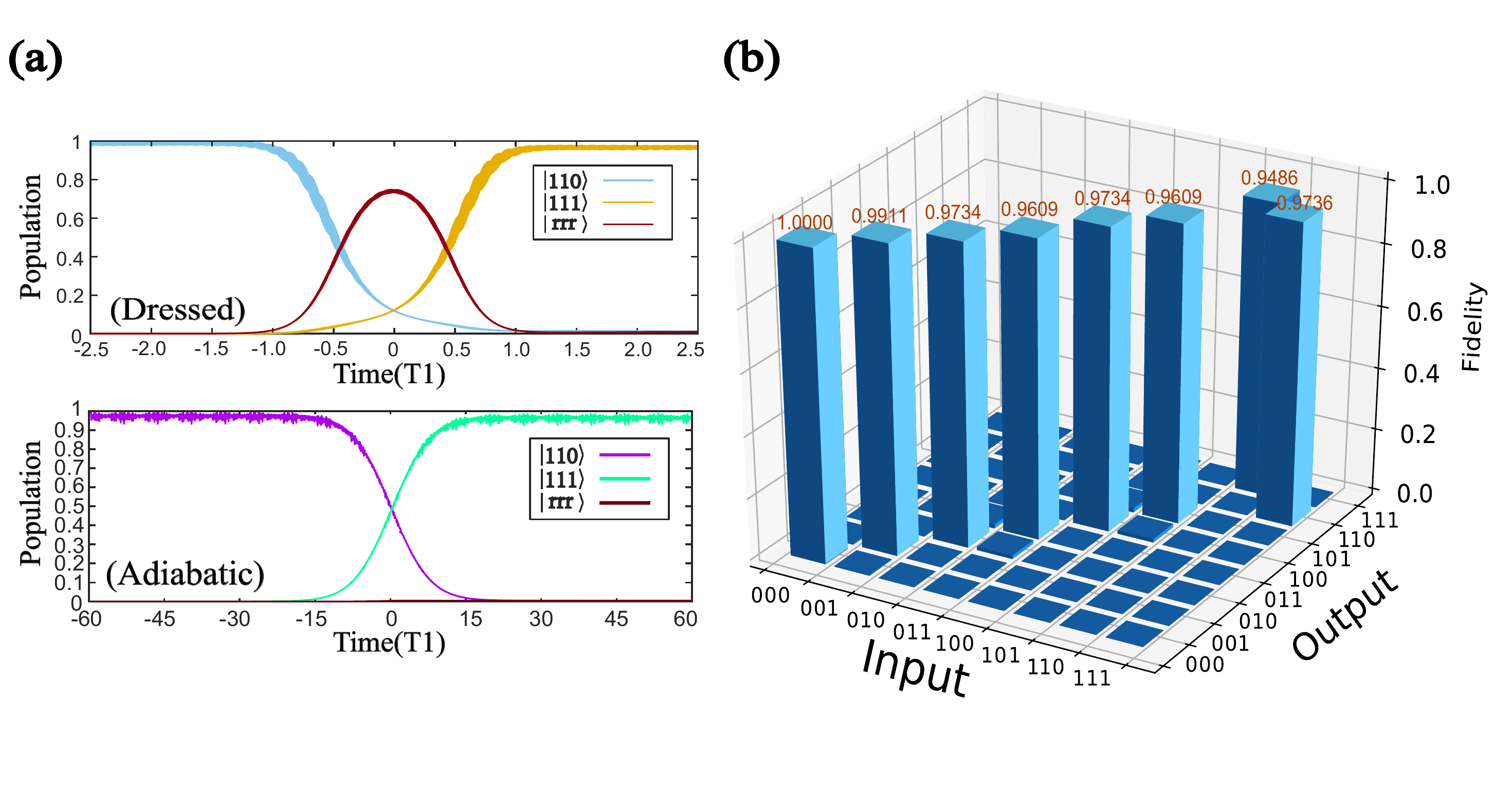}
\caption{The time scale was chosen as T1.  \textbf{(a)} \textbf{Top}: State transfer via Dressed-State Pulses in three qubit gate.The parameters of imposed laser pulse are set as $\Omega/2\pi=30$MHz, $\Omega_c=\Omega$, $\Delta=10\Omega$, $\tau = 0.2/\Omega$. 
\textbf{Bottom}: State transfer via Adiabatic Pulses in three qubit gate.
The parameters chosen: $\Omega/2\pi=30$MHz, $\Omega_c=\Omega$, $\Delta=20\Omega$ and $\tau = 4/\Omega$.
{\textbf{(b)} Ture value diagram of the Toffoli gate with different input states. }   
}
\label{fig:4}
\end{figure}

As a major error source, the spontaneous emissions decreasing the fidelity of quantum logic operation by perturbing evolution of quantum states, can be described by the Lindblad master equation as
\begin{equation}
\begin{array}{rcl}
\dot{\rho}  & = & i \left[\rho, \hat{H}\right] + \sum\limits_{i=1}^{N-1} \sum\limits_{j \in \left\{ 0,1 \right\} }  \mathcal{D}[ L_{ij}]\rho+\sum\limits_{k \in \left\{ 0,1,m \right\}} \mathcal{D}[L_{Nk}]\rho \notag
\end{array}
\end{equation}
where $\rho$ is the density matrix of the quantum system and the dissipations superoperator are defined as $\mathcal{D}[L_{ij}]\rho=L_{ij} \rho L^\dag_{ij} - \frac{1}{2} \left\{ L^\dag_{ij} L_{ij}, \rho \right\} $ with the quantum jump operators $L_{ij} = \sqrt{\gamma_{ij}} \ket{j}_i\bra{r}$ and $\Gamma_{Nk} = \sqrt{\gamma_{Nk}} \ket{k}_N \bra{r}$. For simplicity, we assume $\gamma_{ij} = \gamma/2$, $\gamma_{Nk} = \gamma/3$ and take the spontaneous-emission rate $\gamma$ of Rydberg state $\ket{r}$ as $\gamma/2\pi \simeq 1$ kHz \cite{beterov2009pra}.
For the two-qubit C-NOT gate, the top and bottom in Fig. \ref{fig:2}(d) shows the population transfer and the average fidelity of the gate with the decoherence, respectively.

For the three-qubit Toffoli gate, the value of $V$ is not necessary to be concretely given since we could eliminate it by changing $\Delta$ and $\Omega_c$. To characterize the performance of this three-qubit Toffoli gate by the truth tables with different input states $\ket{\psi_{\mathrm{in}}}$, we calculate the population $|\bra{\psi_{\mathrm{out}}}U(t_f)\ket{\psi_{\mathrm{in}} } |^2$ of all the computational basis states corresponding to each input computational basis state, which forms three $8 \times 8$ truth tables. In Fig. \ref{fig:4}(b), we show the fidelity of corresponding input states. The average fidelity of a quantum gate can be calculated by $F_{\mathrm{av}} = \frac{1}{2\pi} \int_{-\pi}^{\pi} d\theta \bra{\psi_{\mathrm{in}}(\theta)}U^\dag(t_f) \rho(t_f) U(t_f) \ket{\psi_{\mathrm{in}}(\theta)}$ with the input states defined as $\ket{\psi_{\mathrm{in}}(\theta)} = \cos{\theta/2} \ket{110} + \sin{\theta/2} \ket{111}$. The results of numerical simulation in Fig.\ref{fig:4}(b) show that the average fidelity of the three-qubit controlled-NOT gate can reach 95.86\%.

\section{CONCLUSIONS}
\label{conclusions}
In conclusion, our scheme to construct C-NOT gate is presented based on dressed-state method in Rydberg system using Vitanov-style pulses. We gave both equivalent model and initial model as numerical simulation. The result shows that they match well under large detuning condition, robust against parameter fluctuations and decay than other adiabatic case. We also generalize our strategy to multi-qubit cases, by choosing appropriate parameters, $n$-bit Toffli gate could be achieved.
~\\

\section{ACKNOWLEDGEMENTS}
This work was supported by National Natural Science Foundation of China (NSFC) under Grant No. 11804308, No. 12074346 and No. 11804375 and China Postdoctoral Science Foundation (CPSF) under Grant No. 2018T110735
and Natural Science
Foundation of Henan Province (202300410481); Strategic Priority Research Program of the Chinese Academy of
Sciences (XDB21010100); Key R$\&$D Project of Guangdong Province (2020B0303300001).

\appendix

\section{Elimination of off-diagonal elements}
\label{appendix}

The full Hamiltonian $\hat{H}_{\mathrm{new}}$ under dressed-state basis is obtained as \begin{widetext} 
\begin{equation}
\begin{array}{rcl}
\hat{H}_{\mathrm{new}} &=& \left( g_x \sin{\mu}\sin{\xi} - \dot{\eta} - \dot{\theta} \cos{\xi}\sin{\mu} + (\Omega + g_z - \dot{\xi})\cos{\mu} \right) \hat{\sigma}_z\\
& + &\frac{e^{i \eta}}{\sqrt{2}} \left( -i g_x \cos\mu \sin\xi - g_x \cos\xi + i \sin\mu \left(g_z - \dot{\xi} + \Omega \right) + \dot{\theta} \left( -\sin\xi + i \cos\mu \cos \xi \right) + \dot{\mu} \right) \ket{\phi_+}\bra{\phi_0} \\
& + &\frac{e^{-i (\eta + \xi)} }{2 \sqrt{2}} \left(-e^{2 i \xi } (\cos\mu-\Omega)(g_x - i \dot{\theta}) + (\cos \mu + \Omega)(g_x + i \dot{\theta}) + 2 i e^{i \xi } (\sin \mu (g_z - \dot{\xi} + \Omega ) + i \dot{\mu})\right) \ket{\phi_-}\bra{\phi_0} + \rm{H.c.}
\end{array}
\end{equation}
\end{widetext} 
By solving the equations to eliminate all the off-diagonal elements, we can get
\begin{equation}
\begin{array}{rcl}
g_x & = & \frac{\dot{\mu} }{\cos{\xi}} - \dot{\theta} \tan{\xi} \\
g_z & = & -\Omega + \dot{\xi} + \frac{ \dot{\mu}\sin{\xi} - \dot{\theta} }{\tan{\mu} \cos{\xi}}
\end{array}
\end{equation}

\bibliographystyle{apsrev4-1}
\bibliography{ref}

\end{document}